# Is Intersubjectivity Proven? A Reply to Khrennikov and to QBists

**Hervé Zwirn**

Centre Borelli (ENS Paris – Saclay, 4, avenue des Sciences, 91190 Gif-sur-Yvette)
&
IHPST (CNRS, ENS Ulm, University Paris 1, 13 rue du Four, 75006 Paris, France)

herve.zwirn@gmail.com
ORCID : 0000-0001-9299-4281

**Abstract:** In two recent papers Khrennikov [1, 2] uses what he calls Ozawa's intersubjectivity theorem [3] to claim that intersubjectivity is necessarily verified in quantum mechanics and to criticize QBism and more generally all interpretations that are perspectival. In agreement with two previous QBist's papers [4, 5], I explain here why Khrennikov's proof is not valid but in contrast with one of these papers [5], I criticize the way intersubjectivity is dealt with in QBism.

## 1 Introduction

The question of intersubjectivity is a very profound and debated one in quantum physics. In some sense intersubjectivity could be opposed to perspectivalism as described by Leifer [6]. In two recent papers Khrennikov [1, 2] uses what he calls Ozawa's intersubjectivity theorem [3] to claim that intersubjectivity is necessarily verified in quantum mechanics.

I will first show that the so called "Ozawa's intersubjectivity theorem" is a direct consequence of a standard EPR situation and that nothing really new is proven even if it is true that the emphasis on intersubjectivity is not usually what is done when discussing the EPR experiment. I will then argue that analysing what happens in EPR correlations requires to make explicit the interpretational framework in which we situate ourselves. Then I will support the QBist's claim that the conclusion of this analysis inside QBism is not that intersubjectivity should hold between agents. I will make clear that it is neither the case in Convivial Solipsism where the very concept of intersubjectivity has a meaning that is not the standard one. I will also question the way intersubjectivity is dealt with in QBism as expressed in Schack's paper [5] and will compare intersubjectivity and perspectivalism in QBism and Convivial Solipsism in direct line with the analysis provided in Zwirn [7].

## 2   Ozawa's proof

The framework inside which Ozawa proves his result is based on the following situation: A system S interacts at time 0 with two apparatuses intended to measure the same observable A of the system. $M_1$ and $M_2$ are the observables of the apparatuses that observers I and II measure at time $t_1 > 0$ and $t_2 > 0$ respectively. At time 0, the system S is in a state $|\psi\rangle$ and the environment E constituted by the two apparatuses in the state $|\xi\rangle$.

Ozawa uses the Heisenberg representation. If U(t) is the evolution operator of the total system S+E, then $M_1(t) = U(t)^\dagger (I \otimes M_1) U(t)$ and idem for $M_2(t)$. He then makes two reasonable assumptions. The probability reproducibility assumption requires that the probabilities to get the result $x$ during a measurement of the observables $M_1$, $M_2$ and A should be the same. This is indeed what we are looking for from correct measurement apparatuses intended to measure the observable A. The locality assumption assumes that $M_1(t_1)$ and $M_2(t_2)$ commute because the measurements from observers I and II are space-like separated. Then Ozawa shows in a few lines derivation that: $\Pr\{M_1(t_1)= x, M_2(t_1)= y\} = \delta_{xy}$. The mathematics are correct but what they mean is not what Ozawa and Khrennikov claim.

To make the situation more concrete we can assume, without any loss of generality, that the system S is a spin ½ particle and that the observable A to be measured is the z-spin. Let's assume that two Stern and Gerlach apparatuses designed for measuring z-spin interact at time 0 with the particle[1]. Let's furthermore assume that these two apparatuses are sent far away from each other and that observers I and II measure their apparatus at time $t_1$ and $t_2$ such that these two measurements are space-like separated. For the sake of simplicity, we will use the standard way to describe the situation in the Schrödinger representation. Let's $|+\rangle$ and $|-\rangle$ be the eigenstates of A and $|M_1^+\rangle$, $|M_1^-\rangle$, (resp. $|M_2^+\rangle$, $|M_2^-\rangle$) the corresponding eigenstates of the observable $M_1$ of the apparatus I (resp. the apparatus II). At time 0 before the interaction between the system and the apparatuses, the system is in a certain state $\psi = \alpha|+\rangle + \beta|-\rangle$ and the apparatuses in the states $\xi_1$ and $\xi_2$. After the interaction, due to the unitary evolution, the global system is in the entangled state $\psi = \alpha|+\rangle|M_1^+\rangle|M_2^+\rangle + \beta|-\rangle|M_1^-\rangle|M_2^-\rangle$. If we trace over the system S, the reduced density matrix of the two apparatuses will contain the two terms $|M_1^+\rangle|M_2^+\rangle$ and $|M_1^-\rangle|M_2^-\rangle$. We recognize here a typical EPR situation for which, as is well known, any two measurements of $M_1$ and $M_2$ (even space-like separated) on the two subsystems will always give the same results (either ++ or --).

This result is already well known and "Ozawa's theorem" which basically says nothing more, does not bring anything really new except that it is used for defending intersubjectivity. Of course this raises the question of understanding what this result means in term of interpretation. This is here that it is naïve to draw an immediate conclusion according to which this is a proof that quantum mechanics leads necessarily to intersubjectivity and that two observers must necessarily agree in the usual sense of "agreeing". The EPR situation is subtle and even the more complex property of non-locality derived from the violation of Bell's inequality can be discussed and accepted or rejected depending on the interpretation that is chosen. Only a superficial analysis adopting implicitly a standard realist view can reach the conclusion that non-locality is unavoidable. A fortiori this is the case of the property of intersubjectivity which is correct in a simple realist framework but debatable in other framework. Let's now switch to the impact of choosing different interpretations.

---
[1] Let's forget here the fact that a spin measurement is a destructive measurement and that in practice this would be impossible. The argument is intended to illustrate what could happen with two measurement apparatuses interacting with a system as postulated by Ozawa.

## 3 Analysis of EPR type situations
### 3.1 Non-locality

The typical EPR situation is the one in which a pair (U, V) of spin ½ particles is in a singlet state $|\psi\rangle = \frac{1}{\sqrt{2}}\left[\,|+\rangle^U |-\rangle^V - |-\rangle^U |+\rangle^V\,\right]$ and separates. Then measurements of z-spin on U and V are done. It is well known that if one gets + for the measurement on U then one will get – for the measurement on V (and vice versa). In a standard realist framework, where the wave-function is assumed to represent the real physical state of the system, this is often interpreted as meaning that the measurement on U "causes" an immediate effect on the state of V. This notion of immediateness comes from the fact that this is true even if the measurements are space-like separated. Of course, the spontaneous explanation of this strange behaviour is to assume that the values are not created during the measurements but where already determined when the two particles separated. But this is forbidden by the violation of Bell's inequality. So, in the standard realist framework, there is no other choice than to accept the fact that the measurement on U has an instantaneous effect on V (which is called non-locality). This is nevertheless not very satisfying because independently of the weirdness of the thing (Einstein famously called it a spooky action at a distance), it is not possible to determine which measurement is the first one if they are space-like separated. This is nevertheless a point widely accepted in the community of physicists who accommodate themselves with it since it is not possible to transmit information by this process (which is a poor consolation).

Now it is possible to challenge the concept of non-locality if we refuse to adopt a standard realist interpretation. This is the case of QBism [8, 9, 10, 11, 12, 13, 14, 15] which considers that the wave-function is not representing the real state of the system but is a tool allowing an agent to bet on the result she will get if she does a measurement and which moreover says that a measurement is a personal action of an agent on the system and is not directly shared with other agents. Then non-locality is no more required since the results on U and V must be obtained by the same agent and one agent cannot do space-like separated measurements. Moreover, even if after having done the first measurement the agent will bet with probability 1 on the result she will get on the second measurement, for QBists that does not mean that the state of the second system has changed in such a way that it corresponds to this very result. QBists refuse the famous Einstein's criterion of reality according to which:

*"If, without in any way disturbing a system, we can predict with certainty (i.e., with probability equal to unity) the value of a physical quantity, then there exists an element of reality corresponding to that quantity."*

Convivial Solipsism (ConSol) [7, 16, 17, 18, 19, 20, 21, 22] also get rid of any concept of non-locality for similar reasons. Inside ConSol the wave-function is also relative to one observer. A measurement is the perception of a result by an observer while there is no physical collapse of the state. There is only a perception of one of the components of the superposed wave-function. Each observer builds her own reality (which is called the phenomenal reality) through the results she gets by her measurements. For a given observer there is no direct access to the perceptions of another one. The other observers are on the same foot than any other physical system and are described by a possibly superposed wave-function. Asking another observer what she saw amounts to make a measurement on this observer. So the result of a measurement is strictly relative to the observer having done it and is not shared with other observers. Moreover if an observer does a measurement and gets a result this is not considered as a measurement by another observer who considers that the measurement made by the other is just an interaction between the other observer and the system, which leads to an entanglement between the other observer and the system. So there is no sense in which the measurement on U causes an effect on V. First, a measurement is not a physical action changing the state of the system and second, it is only when measuring the observer having done the measurement on V that the first observer perceives the result on V (from which he is not entitled to infer that the result was already there before). Of course this can only be done inside a time-like interval (as for QBism).

This discussion shows clearly that the conclusion about non-locality crucially depends on the choice of one interpretational framework. Non-locality seems to be necessary only inside a realist framework and, by the way, it is rising many issues that cast serious doubts on the coherence of this very framework.

## 3.2 Intersubjectivity

This preliminary discussion was important because the problem of intersubjectivity deserves the same level of precision in the way it is discussed. So let's come back to Ozawa's proof and the use that Krennikov makes of it.

As we mentioned above, it's easy to see that the situation described by Osawa is similar to the situation we used about the measurement of a spin ½ particle. At time 0 before the interaction between the system and the apparatuses, the system is in a certain state $\psi = \alpha|+\rangle + \beta|-\rangle$ and the apparatuses in the states $\xi_1$ and $\xi_2$. After the interaction, due to the unitary evolution, the global system is in the entangled state $\psi = \alpha|+\rangle|M_1^+\rangle|M_2^+\rangle + \beta|-\rangle|M_1^-\rangle|M_2^-\rangle$. The limitation to two dimensions does not change the generality of the result and is simpler for the discussion. Assume that the two apparatuses separate and that a measurement is made on each one.

The two conditions assumed by Ozawa are of course satisfied:

- Locality: $\Pr[M_1 = x, M_2 = y] = \langle\psi'||M_1^x\rangle\langle M_1^x||M_2^y\rangle\langle M_2^y||\psi'\rangle$ for x,y ∈ {+, -} where $|M_1^x\rangle$ and $|M_2^y\rangle$, are the corresponding eigenstates of the observable $M_1$ of the apparatus I (resp. the observable $M_2$ of the apparatus II).
- Probability reproducibility: $\Pr[M_1 = x] = \Pr[M_2 = x] = \Pr[A = x]$

Then it is trivial to see that $\Pr[M_1 = x, M_2 = y] = \delta_{xy}$. The question is now to decide what that means.

We can first notice that the situation described here is exactly similar to the EPR situation analysed above. The only difference is that in the usual EPR experiment the two distant systems that are measured are in a singlet state so that the results obtained are opposite while in this situation the entangled state of the two apparatuses leads to results that are necessarily the same.

We can reproduce the very same analysis here than for non-locality. Inside a standard realist framework where we assume that the wave-function describes the objective state of the systems and so, is the same for all the observers, it is true that if the two measurements are done by two different observers I and II, they will necessarily agree. So intersubjectivity is proven in this framework. Of course, as Khrennikov and Schack notice, each on his side, it is very strange that two random draws having the same probability distribution provides always the same outcomes. Khrennikov [2] says:

> *Why does coincidence of probability distributions imply coincidence of individual outcomes?*

Indeed that would be very surprising if the coincidence of probability distributions of two independent systems would imply the coincidence of individual outcomes but this is not the case here. The two systems are entangled. So there is nothing new if we remind what happens in any EPR situation. It is just another way to see entanglement which is of course a very strange thing in a realist framework!

From the fact that the two meters will give the same outcome Ozawa draws the conclusion that it is a proof that the usual quantum mechanics view according to which a measurement does not ascertain the pre-existing value of the observable is wrong. He claims on the contrary to have proven that a measurement satisfying the condition of probability reproductibility reveals the value the observable has before the measurement. We cannot help thinking of the very early discussion about the EPR experiment and the famous Einstein's criterium of reality. Ozawa's claim is exactly similar to Einstein's claim inside the first version of the EPR argument using position and momentum. Of course this is a priori appealing but it does not work as has been proved more that forty years ago! This amounts to coming back to local hidden variables which are forbidden by the violation of Bell's inequality. So Ozawa's claim [3] against Schrödinger is wrong:

> *Schrödinger argued that a measurement does not ascertain the pre-existing value of the observable and is only required to be repeatable. Since the inception of quantum mechanics, this view has long been supported as one of the fundamental tenets of quantum mechanics. In contrast, we have shown that any probability reproducible measurement indeed ascertain the value that the observable has.*

## 4   Intersubjectivity in QBism

As we emphasized above, discussing the consequences of a mathematical proof requires to make explicit the interpretational framework in which we situate ourselves. Inside a standard realist framework it is fair to recognize that intersubjectivity between observers will be respected. This is not a big discovery as every realist physicist will consider this as obvious. More interestingly, this rises all the difficulties about non-locality (which Ozawa does not even mention while it is at the heart of what he tries to prove) and the strangeness of a concept which has not been really clarified yet. As we said before, it is not enough to be content with the fact that no information can be transmitted…

Inside QBism, Ozawa's argument doesn't stand up to scrutiny as it is clearly showed by Stacey [4] and Schack [5]. For QBists, a quantum state does not represent the physical state of a system but is the agent's encoding of her own personal expectations for what she might experience as a consequence of her actions. Moreover a measurement outcome is a personal experience specific to the agent who does it. As Stacey [4] says:

> *Whether the equations or any of the quantities within them can actually refer to a pair of "two remote observers" measuring "the same observable" is an assumption dependent upon one's interpretation of quantum theory. In QBism, that is simply not what the equation can be talking about.*

What Ozawa calculus proves is simply that an agent will, with probability 1, get the same results if she does a measurement on two apparatuses designed to measure the same observable and having interacted previously with a system.  I also agree with Stacey when he says that taking into consideration the difference between PVM and POVM is irrelevant to the subject.

Now, the interesting question is: "Is intersubjectivity in the usual sense, which is that two agents "really" can share the same experience, possible in QBism or not?". I analyse this question in a recent paper [7] and my conclusion is that QBists should not attempt to preserve intersubjectivity contrarily to Schack's wish [5]:

> *In QBism, intersubjective agreement is not an automatic consequence of the quantum formalism, but a goal that agents might strive for.*

QBism is based on two key assumptions.  The first one is that quantum formalism is nothing more than a tool enabling each agent to make probabilistic bets on the results she will obtain if she performs a given experiment. This formalism does not describe the world or tell us anything directly about it. Such a hypothesis could perfectly well be made about Newtonian mechanics in a classical mode, what we might call CBism. Ontologically speaking, the world would be as it appears to be (standard naive realism), an experiment would be what is usually called an experiment and the results would be objective and shareable with any other experimenter. The world would therefore conform to the standard realist conception, but the mathematical formalism would not be considered as describing anything about the structure of the world, but as a simple artifice for calculating the results that we will obtain if we carry out this or that experiment (the fact that it is not probabilistic here is of no importance in underlining the spirit of this hypothesis). It is an instrumentalist postulate and does not affect the ontology of the world.

The second one is much more involved and is ontological in nature. It is that when an observer carries out an experiment, he obtains a result that 1) does not pre-exist but is created by him, 2) is personal to him. It is this second postulate that makes QBism interesting and distinguishes it from simple instrumentalism.

It is important to realise that although QBism requires the conjunction of these two postulates, they are independent of each other. The first postulate poses no problem and can be accepted as it stands by any physicist who is not anxious to preserve a strict scientific realism. The second postulate, on the other hand, needs to be examined in all its consequences, for they are more engaging than they may at first appear.

QBists are for the moment stuck in the middle of two positions and they have to choose. The first one is to think that QBism does not tell you anything about the world and that it applies only to the very special cases when an agent makes an experiment on a quantum system. That amounts to thinking that it is just a tool that you can use when you want to know what is probably going to happen if you make such an experiment. It is very useful but only for that purpose. It is like a screwdriver. You absolutely need it when you want to put a screw but it does not tell you anything else (or very very little) about the world. Roughly, that comes to considering that the first postulate is the most important and that you draw no consequence on the communication you make with other agents from the second postulate. The second position is to think that agents must apply quantum mechanics for everything they do in the universe and that what QBism says is universally valid. In this case, you must apply the same principles (i.e. the second postulate) for every single experience that an agent has in her daily life, including when she walks, watches a movie, reads a book or talks to other agents. You have the choice between these two positions but you must stay coherent all along the discussion with your choice. The problem is that very often QBists adopt the latter attitude when they speak of physics and the former one when they discuss the problem of intersubjectivity. This is what I already noticed in a previous paper [7] through a quotation from Mermin [23]:

> *Although I cannot enter your mind to experience your own private perceptions, you can affect my perceptions through language. When I converse with you or read your books and articles in Nature, I plausibly conclude that you are a perceiving being rather like myself, and infer features of your experience. This is how we can arrive at a common understanding of our external worlds, in spite of the privacy of our individual experiences.*

Mermin seems to think that the communication between agents is of a classical nature and speaks as if, beside the actions that agents take to create personal results in quantum experiments as QBism explains, the macroscopic world (language, books, articles in Nature, etc…) was out of the quantum framework. And this contradiction is also visible when we compare two Schack's sentences [5]:

> *When Alice is not considering thought experiments but uses the quantum formalism for guidance, e.g., on how to conduct an experiment in a lab, she will not assign quantum states to other agents.*

and:

> *Alice's questions to Bob should be regarded as actions Alice takes on Bob to elicit answers form him. These answers do not exist prior to the action.*

I fully agree with the second sentence. That means that, even though you are not going to assign an explicit quantum state to another agent (which is of course impossible in practice) nevertheless you have to reason as if asking the other agent was exactly identical to an experiment on a quantum system for which (at least sometimes) you can assign such a quantum state. The first Schack's sentence should fairly not be interpreted too literally as meaning that other agents are treated classically. In QBism it is necessary to emphasize the difference between "having an assignment of a quantum state" and "making an experiment providing you with a personal experience". This is the difference between the two postulates mentioned above. You can perfectly make an experiment without having assigned any quantum state. The only difference is that if you have a quantum state then you can make a bet on the result while you cannot if you have no state. But the very action that you do (without assignment of a quantum state) on another agent to create the experience of the answer you hear is of the very same essence than the experiment (with assignment of a quantum state) you do to make a measurement on a quantum system. This is what QBists call (after Wheeler) "a participatory universe". The consequence is that it is not because you do not assign a quantum state to another agent that you can consider that the communication between you and her is of a classical nature, as if everything was happening like in the usual realist way of thinking: "She tells me something that I hear and what I hear is exactly the same thing that she told me and that she is aware to have told me". That is false.

This is a simple logical consequence of the description of QBism given in the FAQ [24] as quoted by Stacey:

> *A "quantum measurement" is an act that an agent performs on the external world. A "quantum state" is an agent's encoding of her own personal expectations for what she might experience as a consequence of her actions. Moreover, each measurement outcome is a personal event, an experience specific to the agent who incites it.*

If we accept the reasonable assumption that the nature of the world is unique whatever the scale and that quantum mechanics is universal, then any act that an agent performs on the external world is what QBism calls a quantum measurement (i.e. there is no act on the external world which are not of this nature). So this is true also for any question that an agent asks to another agent (this is what the second Schack's sentence quoted above says). Then the result of this action, i.e. what the agent hears, is a personal event specific to this very agent. Hence there is no reason to assume that the other agent shares any experience in agreement with the experience of the first agent.

My feeling is that being fuzzy on this point, QBists always oscillate between a pure first person theory (which should be perspectival) and an intersubjective theory allowing to have a shared reality. What I explain in [7] is that if QBists take fully into account the consequence of their assumptions, especially the second postulate, they will be led to a position fully perspectival as ConSol is. Perspectivalism can be contrasted strongly with objectivity (meaning that the results obtained after a measurement, what observers observe, are facts about the universe) but also with intersubjectivity (the results obtained after a measurement are shared by all the observers). In perspectival interpretations, what is true depends on the observer and a result is relative to this observer. There is a distinction between perspectival and relational. Special relativity is relational because two observers may disagree about what they see, for example the elapse of time between two events if they are in different reference frames, but there is nonetheless a one to one transformation between the two points of views and there is a global truth from which the two points of view can be derived. It is similar to seeing a 3D object form different angles. What is seen can be different but there is a global truth from which you can derive what you see. That's not the case in perspectivalism. What is true depends on the observer in a much stronger form than in relationalism.

## 5 Intersubjectivity in Convivial Solipsism (ConSol)

Inside Convivial Solipsism, Ozawa's argument does not work either. I will not here describe ConSol in details and will refer the reader to articles already published where I give a detailed description and compare it with QBism [7, 16, 17, 18, 19, 20, 21, 22)]. It is enough to know that inside ConSol wave-functions are relative to one unique observer (as are also observables or Hamiltonians). This is similar to QBism. A measurement is the perception by the observer of one component of a superposed state. This is what is called "the hanging-on mechanism". The key to understand that is to consider that a superposed state describes "something" that an observer is not able to perceive in all its richness due to the limitations of his brain. So she can perceive only a part of it which is one component. So, as in QBism, a measurement is something that is relative to one observer. A second rule states that once an observer is hung-on to one component (this is done probabilistically in agreement with the Born rule), she will stay hung-on to this branch for all subsequent measurement. This plays the role of the usual reduction postulate but it concerns only the perception and not the system. This is similar to the updating made in QBism when an agent learns a new result.

Other observers are considered as physical systems and the only way for one observer to know what another observer got is to ask her. But asking an observer means measuring her and before the measurement, the second observer can be in a superposed state exactly as any other physical system. Now assume that a first observer makes a measurement of an observable A on a system and that she knows that another observer made the same measurement on the system. For mathematical reasons totally similar to Ozawa's calculus, if the first observer asks to the other what she saw, she will get an answer that is in agreement with the result she got. That is the reason why the name of the interpretation contains the word "convivial". An observer will always agree with another one about the result they got if they did the same measurement. But "conviviality" is not "intersubjectivity". There is no way for an observer to have a direct access to the perceptions of another observer. This is the reason why it is not because Alice hears Bob saying that he got a certain result that Bob "really" got this result. For Alice, Bob is in a state entangled with the system before she asks him. Then she measures him and perceives only one component of Bob's superposed state but, due to the second rule, she can perceive only the component which is in agreement with the result she got herself. But that does not mean that Bob himself perceived the same component when he did his measurement. That can seem very strange but think to the pointer of an apparatus in a superposed state, which is what happens if we assume that quantum mechanics is universal and if the apparatus interacts with a system in a superposed state. The pointer is in a superposed state of positions. Bob is exactly in the same situation from Alice's point of view. The superposition of positions of the pointer is the equivalent for Bob to a superposition of saying that he got different results.  Inside ConSol everything, including sentences, is relative to one unique observer. Each sentence has to be indexed by the observer who is expressing it. There is no way to compare the perception of two observers. So concluding "intersubjectivity" from "conviviality" would be a mistake.

This is the reason why ConSol is maximally perspectival [7]. Each observer builds her own reality (her "phenomenal reality") with the results she gets through measurements on her "empirical reality" which contains all the potentialities that she can possibly actualize. But the phenomenal reality of an observer is outside of any access from other observers. Nevertheless, if an observer makes a measurement on a system and knows (has perceived) that another observer made a similar measurement on the same system, then she can ask the other observer (measure her) about the result she got and the first observer will get an answer in agreement with the result she got on the measurement she made on the system. Notice that this point is not necessary inside QBism which does not make explicit the way an agent gets a result through her action on the world. So it could be possible that two agents disagree on the results they got from the measurement of the same system even if, under suitable conditions of reliability that the first agent attribute to the second, the first agent will bet with probability 1 that they will agree.

# 6    Conclusion

It is true that radical perspectivalism seems very strange and that we would like to find a way to mitigate it. But in a sense that is not something new. Attempts are currently done by several searchers for linking phenomenology and QBism [25, 26]. One of the major problems of phenomenology is how to establish the link between a shared reality and an exclusively first-person discourse. This is not yet a completed task. So, it is not astonishing to see that the same kind of difficulty arises among the most ambitious interpretations of quantum mechanics.

**Funding**: None
**Institutional Review Board statement**: not applicable
**Informed Consent Statement**: not applicable
**Data Availability Statement**: not applicable
**Conflicts of Interest:** The author declares no conflict of interest